# LEEx-B: Low Energy Experimental Bench Development at IPHC-CNRS Strasbourg

E. Bouquerel[a,†], T. Adam[a], C. Maazouzi[a], E. Traykov[a], P. Graehling[a], C. Mathieu[a]

[a] *IPHC, Université de Strasbourg, CNRS-IN2P3, F-67037 Strasbourg, France*

*E-mail*: Elian.Bouquerel@iphc.cnrs.fr

ABSTRACT. As a part of future developments of beam diagnostics, a low energy experimental bench (LEEx-B) has been recently designed, built and commissioned at IPHC-CNRS of Strasbourg. The bench is composed of a $Cs^+$ ion gun installed on a HV platform and providing beams up to 25 keV. A beam profiler and an Allison-type emittance-meter allow the qualification of the setup and also the characterization of the beam. During the commissioning process, the electronics, and the control system were upgraded in order to push the limits towards low beam currents measured by the emittance-meter.

KEYWORDS: Experimental bench; Low beam energy; Beam diagnostics.

# Contents

## 1. Introduction

Beam diagnostics are essential elements in the commissioning, tuning and operation of particle accelerators, transfer beam lines, separators, spectrometers and other beam manipulation systems. Diagnostics are used for measuring beam parameters such as intensity, energy, positioning and distributions of the beam passing through it.

Over the last twenty years many accelerator and accelerator-based projects emerged, which rely on new technologies and deliver beams with extreme energies and intensities for fundamental research, e.g. FCC, LHC, and ESS, whereas others, dedicated to applications and R&D, operate at lower energies but have very specific beam requirements, for the medical sector in particular.

These new developments require measuring certain beam characteristics with increasing precision and reliability and very often without affecting the operation of the machine. However, existing, presently used diagnostics have their limitations and without innovations and expanding the technological boundaries, most of the new accelerator projects would not materialize.

It is within the framework of low energy accelerator developments and applications that the EIA (*Equipe Instrumentation des Accélérateurs*) team at IPHC-CNRS Strasbourg has decided to concentrate a large part of its activities. With nearly 20 years of experience in the field of development of beam diagnostics and control systems of [1-4], the team has devoted itself in the conception and development of future generation diagnostics and improvement of existing systems. One of its major accomplishments is the development of Allison-type emittance-meters. Its principle of operation is based on the design proposed by P. W. Allison et al. in 1983 of a scanner capable of measuring the beam angle distribution with an electric sweep while a mechanical slit scan probes the particles position distribution [5]. Several versions of this device have been developed within projects such as SPIRAL2, MYRRHA, FAIR and soon NEWGAIN [6-9].

Until recently, the EIA team depended a lot on other teams in order to be able to develop diagnostics and more precisely concerning the qualification and validation of the devices. These important steps were often subject to strong geographical (going to laboratories) or temporal constraints (available slots to use the facilities). In order to allow the team to design diagnostics while being free from these constraints and gaining autonomy and efficiency, it was decided to work on developing an experimental bench specific to the team.

## 2. The Low Energy Experimental Bench

Having a low energy experimental bench is interesting for various reasons. It first allows researchers to conduct experiments with minimal energy consumption. Additionally, such a bench helps optimize resource allocation, as it minimizes energy costs associated with running experiments.

Besides the aim of carrying out and validating new processes in the development of beam diagnostics, the Low Energy Experimental Bench (LEEx-B) could be used for various purposes in scientific research and technological applications:

- Surface Analysis and Characterization: Very low energy ion beams are used for surface analysis techniques such as ion scattering spectroscopy (ISS). This technique provides valuable information about the composition, structure, and properties of surfaces [10].
- Thin Film Deposition: Ion beam deposition (IBD) is a technique that uses low energy ion beams to deposit thin films with precise control over thickness and composition. This method is widely used in the fabrication of semiconductor devices, optical coatings, and magnetic storage media [11].
- Surface Modification and Etching: Low energy ion beams are employed for surface modification and etching processes. Ion beam etching (IBE) is used to pattern and etch materials with high precision, making it valuable in microfabrication and nanotechnology [12].
- Ion Implantation: Very low energy ion beams are utilized in ion implantation processes to introduce dopant atoms into materials, altering their electrical, optical, or mechanical properties. This technique is widely used in semiconductor device fabrication and materials engineering [13].
- Ion Beam Analysis: Low energy ion beams are employed in ion beam analysis techniques such as Rutherford backscattering spectroscopy (RBS) and nuclear reaction analysis (NRA). These techniques provide information about elemental composition, depth profiling, and interface analysis of materials [14].

At last but not least, the bench can be used by students for learning and handling a particle beam acceleration and manipulation.

### 2.1. Specifications and equipment

The LEEx-B should be able to validate crucial points when developing beam diagnostics. The first point concerns the accuracy of the beam diagnostics system (BDS). This involves comparing the measured beam parameters with known reference values or established standards to enable reliable measurements. Additionally to this, the precision of the BDS should also be confirmed to allow consistent and repeatable measurements. This can be achieved by performing multiple measurements under similar conditions and analyzing the statistical variations in the results.

Proper calibration of the beam diagnostics instruments is also important. Validation should include assessing the calibration procedures and allowing that the system provides accurate measurements across its entire operating range. The sensitivity of the BDS needs to be determined to ensure that it can detect and measure small changes or variations in the beam parameters. This is particularly important for applications where high sensitivity is required. Furthermore, the compatibility of the BDS with the specific beam characteristics and experimental setup needs to be defined. This involves ensuring that the system can handle the beam properties, such as energy, intensity, and beam size, without introducing significant distortions or limitations. The reliability of the BDS should be validated through long-term testing and monitoring. This includes assessing its performance over extended periods, evaluating its resistance to environmental factors, and identifying any potential failure modes.
Assessing these points leads to more robust and trustworthy measurements of beam parameters.

The LEEx-B is installed on a modular support with a high voltage platform integrating a low energy ion gun with a surface ionization source and a beamline section that will be composed of optical equipment, diagnostics and a dedicated vacuum chamber intended to accommodate future diagnostics. Alternatively, this chamber can also be used for performing experiments requiring an ion beam.

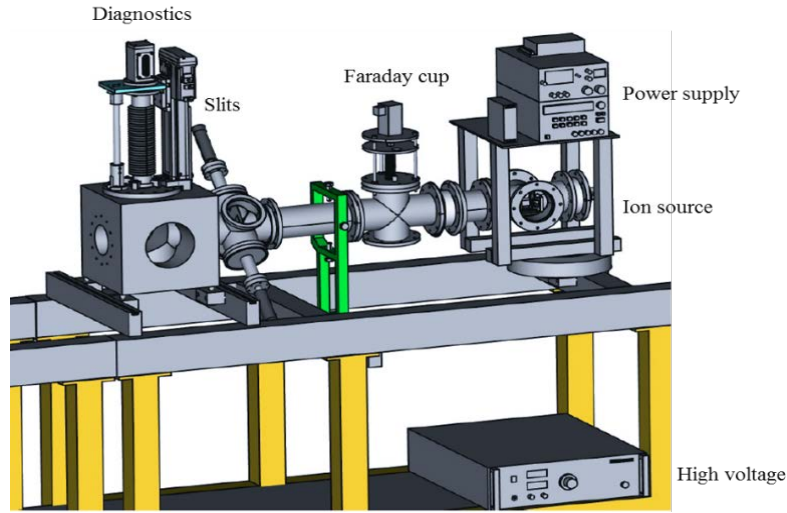

Figure 1. 3D view of the experimental bench.

The beam is generated by a HeatWave HWIG-250 Ion Gun [15] using a cesium compound pellet that enables to produce a beam of 1 µA intensity at 10 keV. The source is installed in a vacuum chamber, itself mounted on a high voltage platform presently limited to 15 kV. The production principle (Fig.2, Fig.3) consists of heating the Cs pellet, thus increasing surface ionization probability, and then forming a beam of the ions via an extraction electrode. The beam passes through a focusing electrode allowing adjustment of its optical properties.

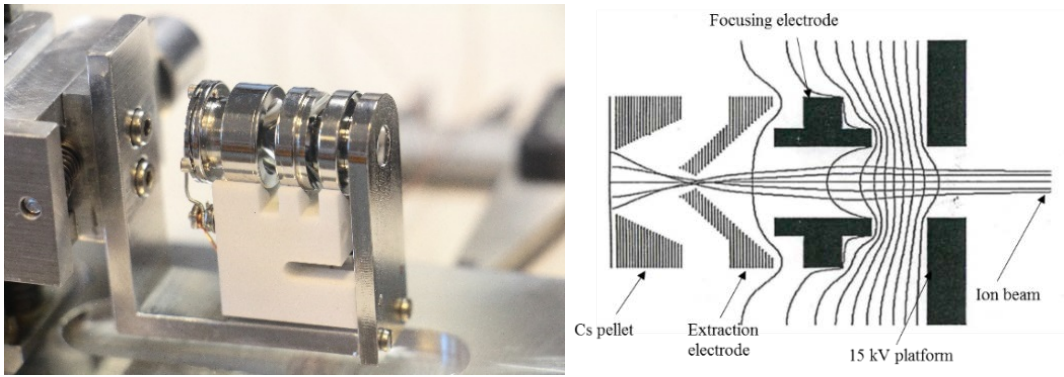

Figure 2. Photograph (left) and principle (right) of the ion gun.

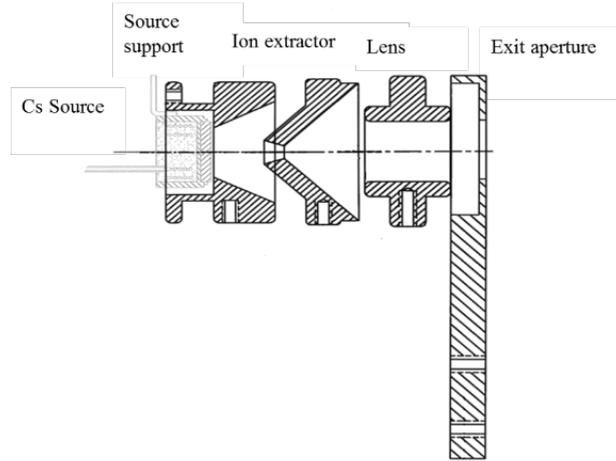

Figure 3. Component layout of the ion source.

The supply of power is divided into three circuits: 1) the "standard" circuit that supplies the components of the test bench excluding the high voltage platform and the safety circuit. 2) the safety circuit that secures the high voltage platform. This circuit is made up of interlocks prohibiting access to the components of the HV platform and a light beacon indicating the state of the source: green light (no risk for intervention in the HT zone); orange light (short-circuit rod out of zone, source power supply active) and red light (interlock active; faraday cage door closed, lock active and short-circuit rod attached) and platform HV power supply on. 3) The mains power supply of the platform (Fig. 4).

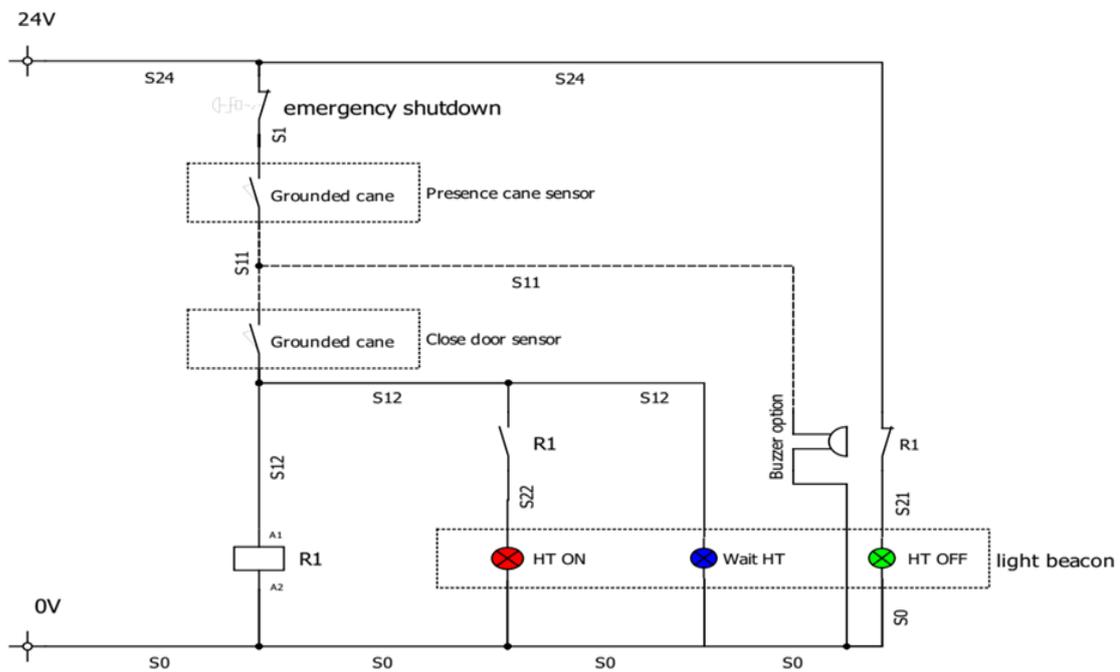

Figure 4. Layout of high voltage platform security.

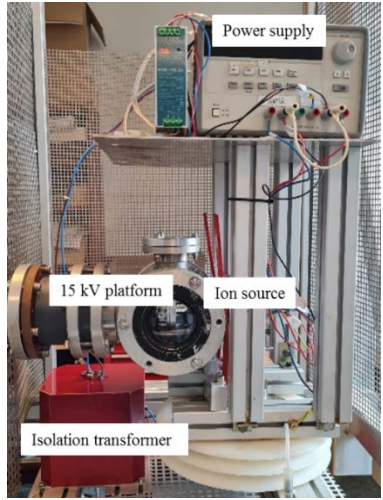

Figure 5. Photograph of the high voltage platform.

The high voltage platform (Fig. 5) is brought to a potential of 15 kV using a FUG HCN 35 power supply, the HT output of which is connected to the ground of the platform. The electrical equipment (HV power supply, +/-24 V power supply, 24 V transformer) is supplied with 220 V via a STANGENES SIT 30-500 "dry" type insulation transformer.

The ion source and its components are maintained under a vacuum of the order of $10^{-6}$-$10^{-7}$ mbar. Two pumps (Edwards nXDS10i Scroll Pump and Edwards nEXT400D turbo-molecular pump) are used to obtain such a level of vacuum. If necessary, additional pumping is possible. The pressure control is done via an Edwards PGC201 gauge box equipped with a Pirani and Penning gauge.

To define the equipment and its positioning necessary for the desired transport of the particles along the bench, optical simulations were carried out taking into account the characteristics of the source (Fig. 6).

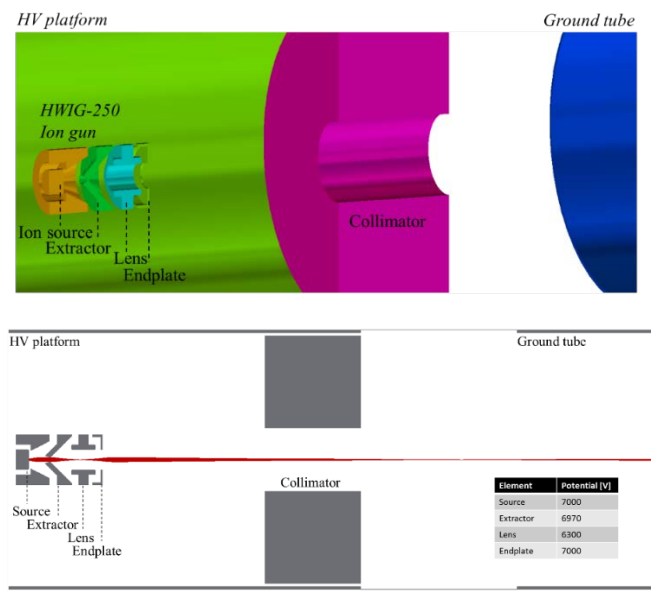

Figure 6. Simulations of the ion gun set up containing the model in SIMION [16] (up) and the trajectories of the ions (bottom).

## 2.2. The control system

The LEEx-B bench must be distinguished according to 3 components: source, optical elements, and diagnostics:
- The source is controlled by four power channels (source, heater, extraction, and focus). These power supplies are placed in the source chamber and therefore at the acceleration potential (beyond 18 kV). The connection with the control and command computer is ensured by an optical fiber insensitive to the 18 kV potential gradient between the two environments.
- As optical elements, two rings (acting as lens) and four plates (acting as electrostatic XY steerer), are controlled by other low voltage power supplies and amplifiers.
- Among the diagnostics, there are two Faraday cups (CF), located at the exit of the source and at the end of the beamline, a wire profiler (PF), and two emittance-meters (EMU), one horizontal and one vertical.

The control architecture is composed of two parts: one part dedicated to the source and a second part dedicated to the optical elements. The whole is transcribed on a single human-machine interface (GUI = Graphical User Interface). Furthermore, the diagnostics are managed by their own GUI, free to use, but taking into account the different inter-diagnostics constraints (interlock). The data collected by these three types of diagnostics in the first instance is a current reading; for the CF, the current read is a translation of the beam current or charge per unit of time at a given location; the use of two CFs, one at the entrance and the other at the end of the line allows to evaluate the beam transfer and adjust the optical elements accordingly; the current reading is done on a Keithley electrometer capable of reading low currents up to picoampere and controlled via the GPIB communication protocol.

The wire profiler, on the other hand, will only give the current induced by the beam on a 0.8 mm diameter wire; by moving the wire at regular or irregular steps, for each position an intensity of current is obtained, after a transverse scan of the beam, we obtain a transverse profile on an axis (X or Y) of the beam. The current is read this time by a pico-femto-ampere meter controlled via TCP/IP.

Finally, the emittance-meter is also based on a current reading. It rather translates the amount of charge diffusing at a given angle. In the end, the so-called transverse emittance information which translates the distribution of (particles) charges according to their angles and the position in the beam is obtained. As the system is more complex, it is managed by a compactRIO. The interaction with the control and command is done via shared variables that allow such a structure to be integrated into an EPICS environment. The GUI interacts with the compactRIO via these shared variables and data processing is done on these remote interfaces at the operator's discretion. Figure 7 shows the layout of the command control of the platform.

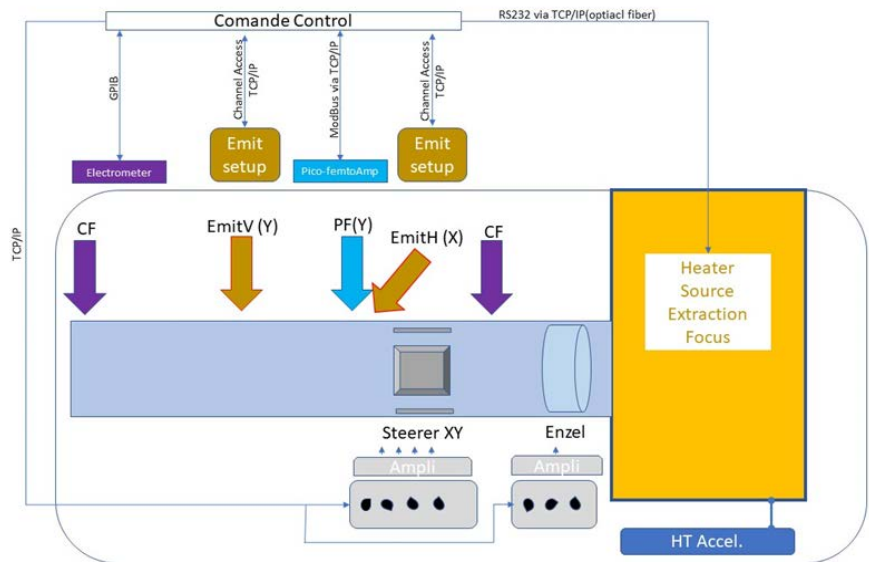

Figure 7. Command control schematic layout.

## 3. Commissioning of the LEEx-B

The qualification of the source was done in two stages. The aim of the first step was to obtain a first result with the test bench and thus validate its concept. It was carried out without an isolation transformer and limited in voltage (3 kV maximum of the platform). The second stage was done with additional optical equipment and a nominal platform voltage of 15 kV.

To generate the ion beam, the high voltage platform is equipped with an Agilent E3631E power supply capable of generating up to 6 V – 5A for heating the Cs pellet and +/-25 V for extracting and controlling a high voltage power supply of source focusing lens. Control and command of the Agilent power supply is done remotely via a private WIFI network. To measure the current at the source output (40 cm from the extraction), a Keithley 6485 pico-ammeter is connected to a Faraday cup, under which an Einzel lens is fixed, to refocus the beam in a second Faraday cup placed 60 cm downstream the first one (Fig. 8).

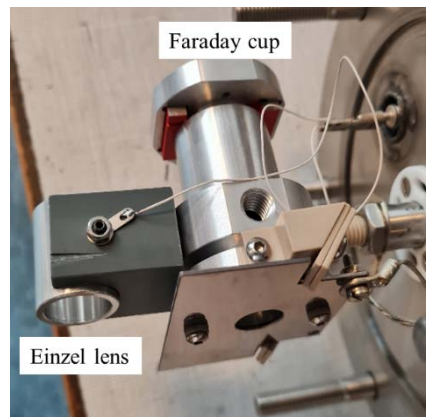

Figure 8. Focusing and current reading.

Different source settings of the source were applied to determine baseline values:
- High voltage platform: 3 kV,
- Heater (pellet heating): 3.2 V and 2.42 A,
- Extraction voltage: -25 V,
- Focusing voltage: 190 V,
- Output current: ~50 nA

### 3.1. Beam profile measurements

Beam profile measurements were done using a grid profiler, currently being developed at IPHC. Figure 9 shows a schematic view of the profiler. It is composed of a 2 x 16 metallic wires of 0.38 mm radius stuck onto a circular frame of 36 mm diameter. The displacement of the head of the grid is motorized by 115 mm. The maximum beam diameter that the diagnostics can measure is 15 mm. The grid is positioned at 1100 mm away from the extraction source.

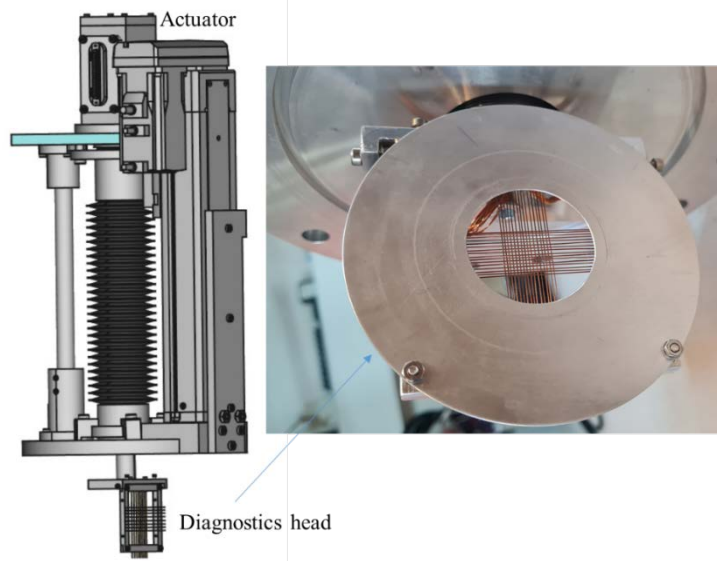

Figure 9. Schematic view of the grid diagnostics (left) and its picture (right).

The first test consisted in varying the voltage of the Einzel lens in order to see the impact on the profile of the beam. Figure 10 shows how the beam profile behaves using Einzel voltage from 0 to 2 kV. Although increasing the lens voltage shows the beam focusing as expected, the results suggest a misalignment of the equipment along the experimental bench. , i.e. a position shift of the Einzel lens relative to the beam axis.

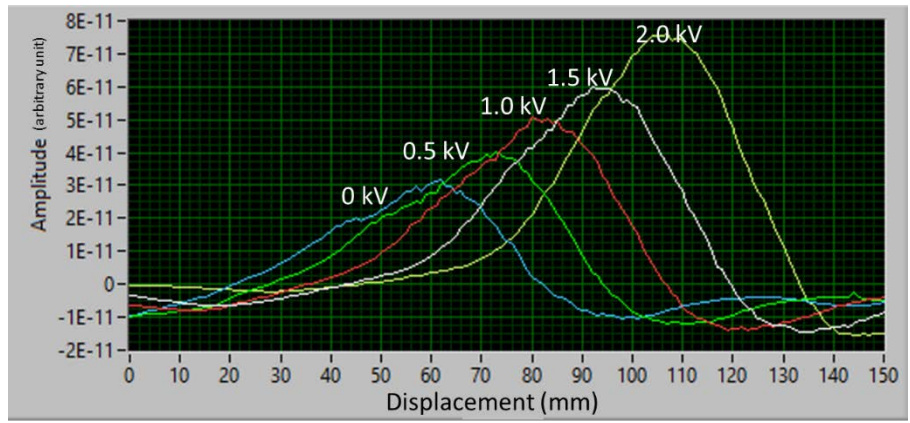

Figure 10. Effect of the voltage of the Einzel lens onto the beam particles (from 0 kV-blue to 2.0 kV-yellow).
The source voltage is kept at 3 kV.

The second test consisted in varying the voltage of the source to see the impact onto the behavior of the particles at the profiler location. The experimental was done when no voltage is applied to the Einzel lens. Results show an increase of the amplitude of the beam when applying from 8 to 15 kV into the source (Fig. 11).

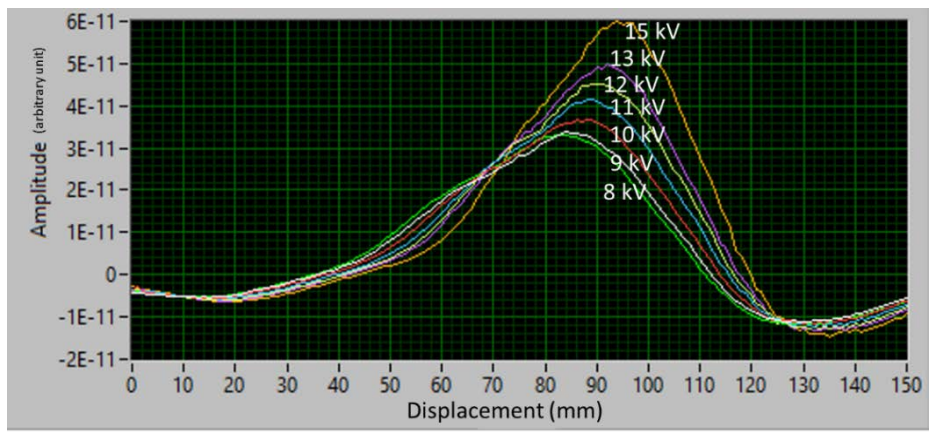

Figure 11. Effect of the voltage of the source onto the beam particles (from 8kV-green to 15kV-orange)

### 3.2. Emittance measurements

Transverse emittance measurements are done using an Allison-type emittance-meter (Fig.12) developed by the EIA team of IPHC [17]. The system is a combination of an electric trajectory sweep technique together with a mechanical position beam sweep. Thus a simultaneous measurement of the position and angle is performed for each position enabling the reconstruction of the full phase space display and calculation. The measurement ranges are from 10 pA to 1 µA for a maximum beam diameter of 80 mm. The minimum possible displacement of the emittance-meter head is 0.1 mm.

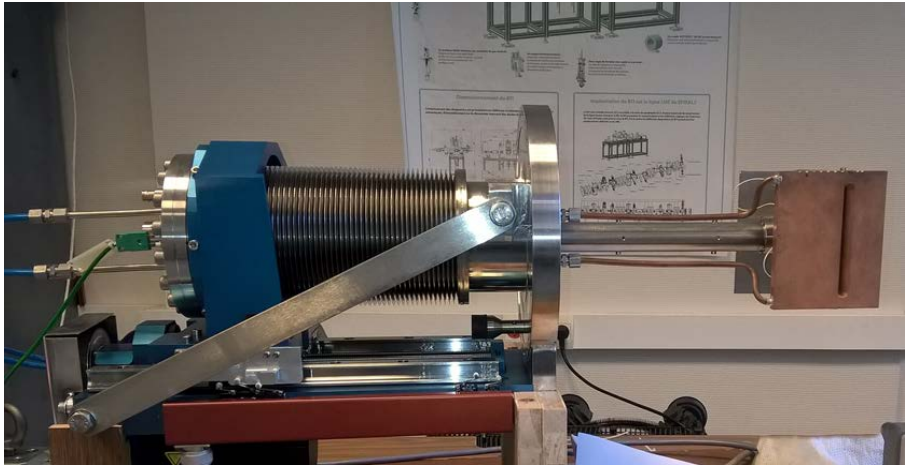
Figure 12. IPHC Emittance-meter.

Figure 13 shows first transverse emittances obtained when varying the Einzel lens. The shape of the measured emittances suggest a possible misalignment in some devices along the LEEx-B.

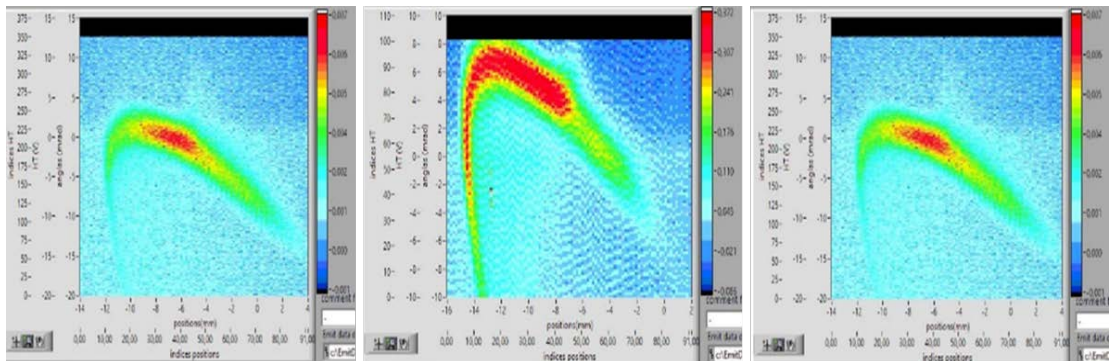
Figure 13. Transverse emittances for different Einzel 1 and 2 settings: 1300 V and 1260 V (left), 1360 V and 1360 V (middle) and 1340 V and 1320 V (right). High voltage platform: 3kV. Extract ion: -25 V. Focus: 190 V. CPU 1, output current: ~30 nA, CPU 2 current: between 15 nA and 24 nA depending on the settings.

### **I/V conversion cards in the measurement of the emittances**

The emittance-meter uses an electronic card placed as close as possible to the diagnostics. This card, named I/V card, enables recovering the current from the latter (Faraday cup, profiler, slits ...) and convert this signal into voltage, in order to limit signal losses and parasitic pick-up contributions. An acquisition card reads back the signal to give the current value. Reading ranges can be adjusted to match the output signal to the input signal (10 fA to 100 µA). The accuracy and reliability of these cards are of great importance in emittance-meter devices. They must be carefully calibrated and characterized to ensure precise measurements. Any inaccuracies or inconsistencies can lead to erroneous emittance calculations, potentially impacting the overall performance of the particle beam system. Furthermore, I/V cards should have a wide dynamic range to accommodate different beam intensities and energies. They should also possess high sensitivity and low noise levels to capture even the smallest variations in the current-voltage characteristics.

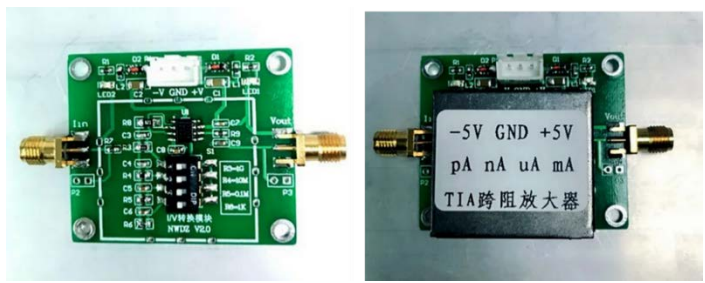

Figure 14. Picture of the front (left) and the back (right) of a I/V card used for the emittance-meter (TLC2201 type).

Figure 14 presents the picture of a card that can be used for converting intensity into voltage for the Allison emittance-meter to operate. The module allows current conversion of up to 50 µA. However, it is possible to work in lower ranges with better resolution thanks to the four controllable gains: 50 nA (Gain 1), 500 nA (Gain 2), 5 µA (Gain 3) and 50µA (Gain 4). The choice of gain (current range) is made via the NI 9477 module. It is also possible to generate two internal currents (10 µA and 100 µA) to test and debug the I/V card and/or the connection with the ADC. The module is equipped with a differential output in the +/- 5 V range.

Once the setup realigned, the LEEx-B allowed characterizing and thus comparing several types of cards in the framework of a new diagnostics currently being developed within the team. Figure 15 shows an example of transverse emittances measured using one of them. The shape of the emittance indicates a better alignment of the components on the bench. Normalized emittance of the beam was found to be 0.057±0.026 π.mm.mrad.

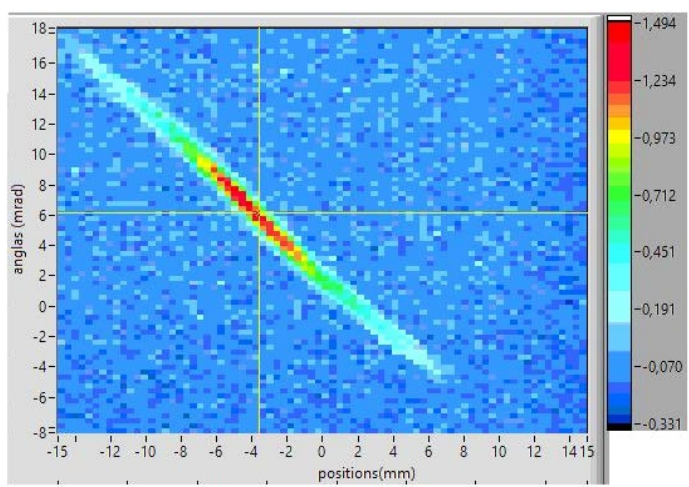

Figure 15. Emittance obtained after re-alignment with one of the I/V card. Accelerating voltage: 25 kV. Beam current on FC: 1.6 nA. Einzel voltage: 5.862V. Steerer voltage: 2160 V.

Besides the fact that the Allison scanner allowed measuring transverse emittances and characterizing the ion source, these results demonstrated for the first time that this device, originally designed for large currents, can measure currents of few nano-Amps. The transmission remains correct, depending on the settings, despite the alignment and the weak acceleration tension.

## 4. Conclusions

A Low-Energy Experimental Bench (LEEx-B) has been developed within the EIA team at IPHC-CNRS of Strasbourg. One of its aims is to provide a dedicated beamline with an integrated $Cs^+$ ion source that allows the team to develop beam diagnostics in complete autonomy from conception to commissioning. The bench has been characterized using a beam profiler and an emittance-meter developed at IPHC. It allowed the comparison between several I/V cards, important for the emittance-meter to work properly. Emittances could be achieved by direct current measurements at very low intensities corresponding to beam currents in the order of 1 nA at the exit of the source.

## 5. Acknowledgments

The authors would like to thank the management of IPHC and CNRS-IN2P3 for trusting and supporting this development.